\documentclass{aa}

\usepackage{graphicx}
\usepackage{txfonts}
\usepackage{booktabs}
\usepackage{xcolor}
\usepackage{hyperref}
\usepackage{natbib}

\newcommand{\Av}{A$_V$}
\newcommand{\jwst}{\textit{JWST}}
\newcommand{\hst}{\textit{HST}}
\newcommand{\halpha}{H$\alpha$}

\defcitealias{sarrouh26}{Sarrouh \& Asada et al. (2026)}

\makeatletter
\AtBeginDocument{\let\linenumbers\relax}
\makeatother

\begin{document}
\nolinenumbers

\title{A New Window on the \halpha\ Luminosity Function and Star Formation Rate Density from $1.2<z<6.6$ from JWST Medium-Band Photometry}
\titlerunning{\halpha\ LF and SFRD at $1.2<z<6.6$ from JWST Medium-bands}
\authorrunning{Martis et al.}

\author{Nicholas S. Martis\inst{1}\corrauth{nicholas.martis@fmf.uni-lj}
\and Chris Willott\inst{2}
\and Gregor Rihtar\v{s}i\v{c}\inst{1}
\and Vesna Pirc Jev\v{s}enak\inst{1}
\and Roberto Abraham\inst{3}
\and Yoshihisa Asada\inst{3}
\and Maru\v{s}a Brada{\v c}\inst{1,4}
\and Gabriel Brammer\inst{5,6}
\and Guillaume Desprez\inst{7}
\and Vicente Estrada-Carpenter\inst{8,9}
\and Kartheik Iyer\inst{10}
\and Lamiya Mowla\inst{11}
\and Adam Muzzin\inst{12}
\and Gaël Noirot\inst{13}
\and Ghassan T. E. Sarrouh\inst{12}
\and Marcin Sawicki\inst{14}
\and Sunna Withers\inst{12}
\and Giordano Felicioni\inst{1}
\and Jon Jude\v{z}\inst{1}
\and Danilo Marchesini\inst{15}
\and Vladan Markov\inst{1}
\and Katherine Myers\inst{12}
\and Luke Robbins\inst{14}
\and Visal Sok\inst{16}
\and Wren Suess\inst{16}
\and Roberta Tripodi\inst{17}
}

\institute{
University of Ljubljana, Faculty of Mathematics and Physics, Jadranska ulica 19, SI-1000 Ljubljana, Slovenia
\and NRC Herzberg, 5071 West Saanich Rd, Victoria, BC V9E 2E7, Canada
\and David A. Dunlap Department of Astronomy and Astrophysics, University of Toronto, 50 St. George Street, Toronto, Ontario, M5S 3H4, Canada
\and Department of Physics and Astronomy, University of California Davis, 1 Shields Avenue, Davis, CA 95616, USA
\and Cosmic Dawn Center (DAWN), Denmark
\and Niels Bohr Institute, University of Copenhagen, Jagtvej 128, DK-2200 Copenhagen N, Denmark
\and Kapteyn Astronomical Institute, University of Groningen, P.O. Box 800, 9700AV Groningen, The Netherlands
\and School of Earth and Space Exploration, Arizona State University, Tempe, AZ 85287, USA
\and Beus Center for Cosmic Foundations, Arizona State University, Tempe, AZ 85287, USA
\and Columbia Astrophysics Laboratory, Columbia University, 550 West 120th Street, New York, NY 10027, USA
\and Whitin Observatory, Department of Physics and Astronomy, Wellesley College, 106 Central Street, Wellesley, MA 02481, USA
\and Department of Physics and Astronomy, York University, 4700 Keele St., Toronto, Ontario, M3J 1P3, Canada
\and Space Telescope Science Institute, 3700 San Martin Drive, Baltimore, Maryland 21218, USA
\and Department of Astronomy and Physics and Institute for Computational Astrophysics,  Saint Mary's University, 923 Robie Street, Halifax, Nova Scotia B3H 3C3, Canada
\and Department of Physics and Astronomy, Tufts University, 574 Boston Ave., Medford, MA 02155, USA
\and Department of Astrophysical and Planetary Sciences, University of Colorado, 2000 Colorado Ave., Boulder, CO 80309, USA
\and INAF -- Osservatorio Astronomico di Roma, Via Frascati 33, Monte Porzio Catone, 00078, Italy
}
\date{}

\abstract
{We present the first self-consistent measurement of the \halpha\ luminosity function over a wide redshift range, covering cosmic noon into the epoch of reionization. Our analysis utilizes a novel method based on \textit{James Webb Space Telescope} (\jwst) NIRCam medium-band imaging. We combine data from the CANUCS, \jwst\ in Technicolor, and JUMPS surveys which offer deep, uniform imaging (29.5-30 AB, $3\sigma$) with extensive NIRCam medium-band coverage, reaching up to 29 total filters (up to 20 \jwst) when including ancillary \textit{Hubble Space Telescope} (\hst) ACS and WFC3/UVIS data. The superb spectral energy distribution (SED) sampling enables precise, reliable photometric redshift estimation (outlier fraction 1.7\%, $\sigma_{NMAD}=0.039$ for this sample) as well as accurate continuum subtraction and line flux measurement verified by spectroscopic follow-up (no systematic offset, 0.23 dex scatter). We measure the \halpha\ luminosity function (LF) from $1.25 < z < 6.6$ by tracing the \halpha\ emission line in 11 medium-band filters. The combination of depth, redshift coverage, and statistical power is unique, providing strong constraints on the shape of the LF over almost three orders of magnitude in luminosity. Our dense SED sampling enables us to reliably correct for dust attenuation and derive dust-corrected star formation rate functions as well as the evolution of the cosmic star formation rate density over the full redshift range. We recover the peak at $z\sim2$ and a decrease toward $z=6$, though with a higher normalization more in line with recent IR measurements than UV, though eclipsing both. This potential tension will be addressed in future work utilizing larger surveys with MIR coverage to better constrain the bright end of the luminosity function and the effects of dust.}


\maketitle

\section{Introduction}
\label{sec:intro}

Tracing the history of star formation across cosmic time is one of the fundamental goals of extragalactic astrophysics. The \halpha\ emission line has long served as a reliable indicator for galaxy star formation rates (SFRs) \citep{kennicutt83, kennicutt98_sf}. Since the line originates from the recombination of hydrogen in HII regions ionized by short-lived, massive stars, it accurately traces star formation on $\sim 10$ Myr timescales \citep{kennicutt98_sf, calzetti13}. Until recently, this tracer has been effectively employed to measure the cosmic star formation rate density (SFRD) up to $z \sim 3$, the limit at which the \halpha\ line redshifts beyond the observing capabilities of ground-based facilities and the \textit{Hubble Space Telescope} \citep[e.g.][]{Ly07, hayes10, lee12, kashino13, sobral13, pirzkal13, steidel14, sobral16, nagaraj23}. 

At higher redshift, the ultraviolet (UV) continuum has traditionally served as the most accessible SFR indicator due to its origination from young, massive stars, allowing measurement of the SFRD into the epoch of reionization \citep{wyder05, reddy09, cucciati12, schenker13, bouwens15, finkelstein15, donnan23, harikane23, donnan24, mcleod24}. Since the rest-frame UV luminosity of high-redshift galaxies can be straightforwardly determined from optical-NIR photometry, and large imaging campaigns with both ground-based observatories and the \textit{Hubble Space Telescope} enabled efficient sampling across a wide range of intrinsic luminosities, these measurements long provided the bulk of our knowledge of star formation at $z>3$. On the other hand, UV measurements suffer from the limitations of sampling a coarser star formation timescale ($\sim 100$ Myr) due to contribution from less massive stars and more severe attenuation by dust \citep{calzetti00, hao11, murphy11, kennicutt12}. 

In principle, the mm/FIR provides a more accurate tracer of the SFRD since it is unaffected by dust obscuration. However, large-scale surveys with FIR instruments such as \textit{Spitzer} and \textit{Herschel} are significantly limited by shallow depths and difficulty in optical/NIR counterpart identification due to the large beam size \citep{burgarella13, schreiber15, wang19}. ALMA has made significant progress in this regard, with blind surveys able to measure the obscured SFRD out to $z \sim 7$ \citep{zavala21, traina24}. The absolute number of sources detected by these surveys, however, remain small, making it difficult to construct large samples covering a range in redshift and luminosity. Future proposed missions including the PRobe far-IR Mission for Astrophysics (PRIMA) stand to make significant advances in this regard. On the other hand, FIR-based SFRs may be less reliable for low-mass galaxies and at high-z due to lower metal/dust content in these populations and uncertain conversions from IR luminosity to SFR without well-sampled FIR spectral energy distributions (SEDs).

Since the launch of the \textit{James Webb Space Telescope} \citep[\textit{JWST}][]{gardner23}, direct spectroscopic measurement of \halpha\ at $z>3$ has become widely accessible for the first time. In particular, wide-field slitless spectroscopy with the NIRCam grism has uncovered thousands of \halpha\ emitters up to $z \sim 6$ through blind surveys which avoid the potential biases of targeted spectroscopic followup \citep{matharu23, lin24Aspire, covelo-paz25, lin25, fu25}. These studies now provide measurements of the \halpha\ luminosity function (LF) from $4.5 < z < 6.5$, covering areas of the order tens of square arcminutes and therefore beginning to sample representative cosmological volumes and avoid large systematic uncertainties due to cosmic variance. On the other hand, grism spectroscopy suffers from challenges such as contamination from other sources, and arguably more complex reduction, extraction, and modeling of the data compared to imaging.

This work follows an alternative strategy, utilizing emission line boosting of medium-band photometry to infer line fluxes. This approach was employed at high redshift by a number of studies utilizing \textit{Spitzer}/IRAC broadband photometry to measure \halpha\ emission \citep{schaerer10, rasappu16, faisst19, asada22, stefanon22, bollo23} and has recently been verified by \jwst\ grism measurements \citep{covelo-paz25, lin25, fu25}. However, due to the wide filter bandpasses and limited sensitivity, most of these measurements were restricted to the bright end of the \halpha\ luminosity function. Now the strategies employing the \jwst\ NIRCam medium-bands can very efficiently select clean samples of emission line galaxies \citep{williams23, withers23, simmonds24c, martis25}. Moreover, emission line fluxes and equivalent widths can be reliably measured from medium-band photometry \citep{withers23, lorenz25}.

We combine data from the programs CAnadian NIRISS Unbiased Cluster Survey (CANUCS), \jwst\ in Technicolor, and the \jwst\ Ultimate Medium-Band Photometric Survey (JUMPS) to generate a survey consisting of deep, densely sampled photometry. Together these provide a large dynamic range in emission line fluxes sampling the faint end of the \halpha\ luminosity function particularly well and enabling a self-consistent study of its evolution from $1.5<z<6.5$ for the first time.

This paper is organized as follows. In Section \ref{sec:data} we present the data utilized in this work. Section \ref{sec:analysis} describes the sample selection and emission line measurement method. Section \ref{sec:results} showcases our luminosity function results. In Section \ref{sec:Discussion} we derive and discuss our measurement of the cosmic star formation rate density. Finally we draw conclusions in Section \ref{sec:Conclusion}. Throughout the paper we assume a standard $\Lambda$CDM cosmology with $\Omega_\Lambda = 0.7$, $\Omega_M = 0.3$, and $H_0 = 70$ km s$^{-1} Mpc^{-1}$ as well as a Chabrier initial mass function \citep{chabrier03}. All magnitudes are in the AB system \citep{oke83}.  

\begin{figure*}
\centering

\begin{minipage}[t]{0.68\textwidth}
    \centering
    \includegraphics[width=\linewidth]{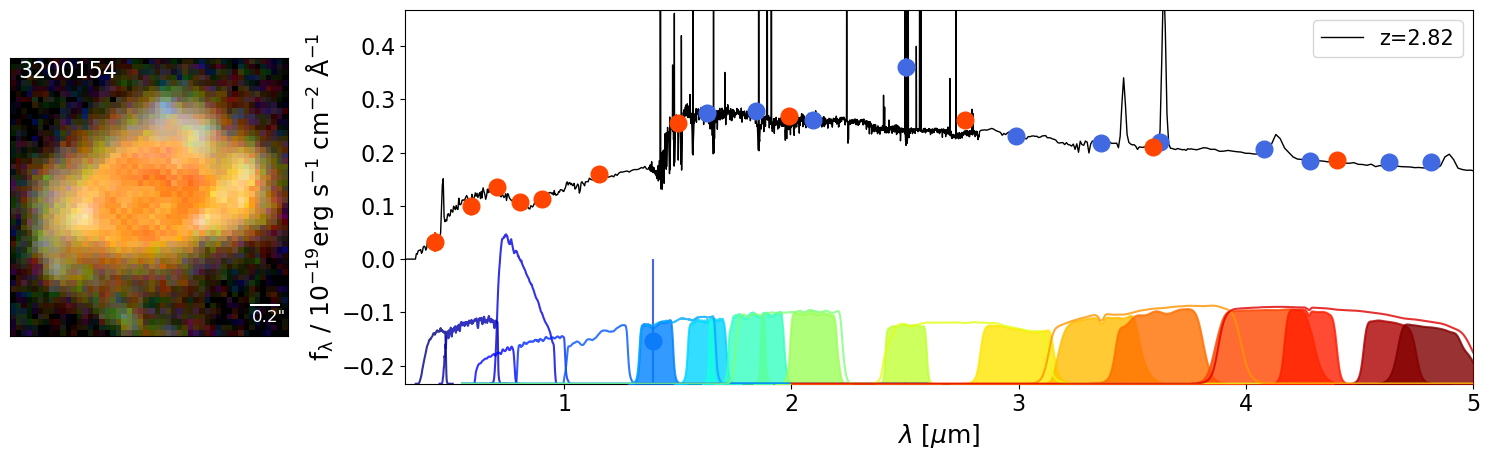}
\end{minipage}
\hfill
\begin{minipage}[t]{0.31\textwidth}
    \centering
    \includegraphics[width=\linewidth]{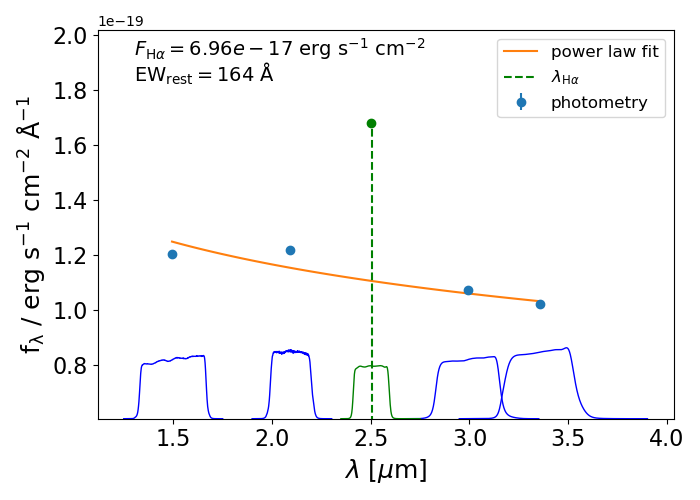}
\end{minipage}


\begin{minipage}[t]{0.68\textwidth}
    \centering
    \includegraphics[width=\linewidth]{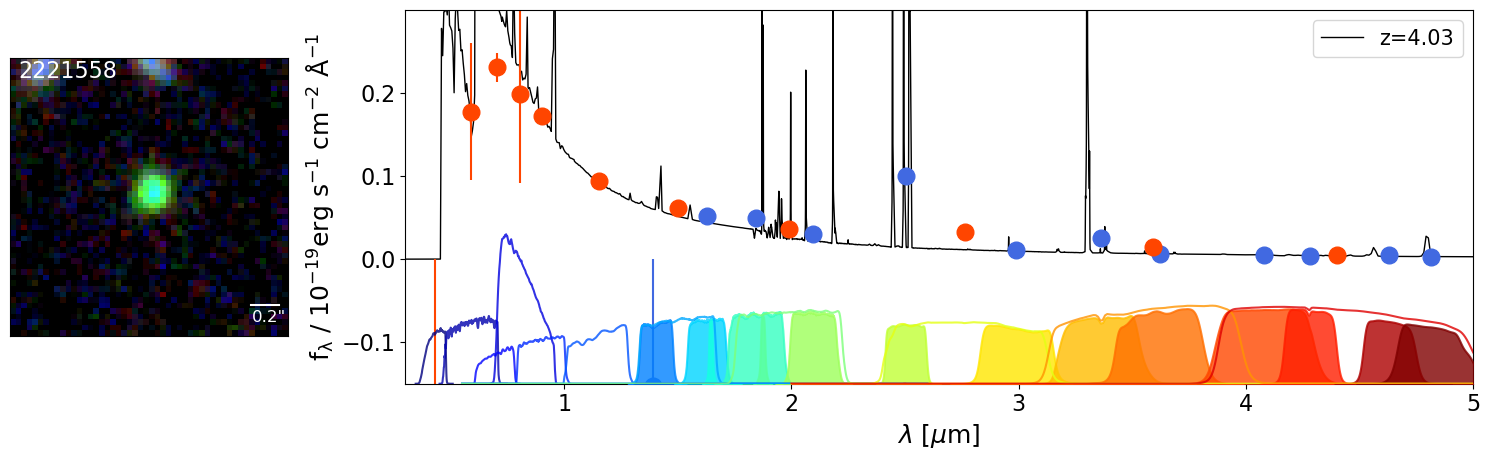}
\end{minipage}
\hfill
\begin{minipage}[t]{0.31\textwidth}
    \centering
    \includegraphics[width=\linewidth]{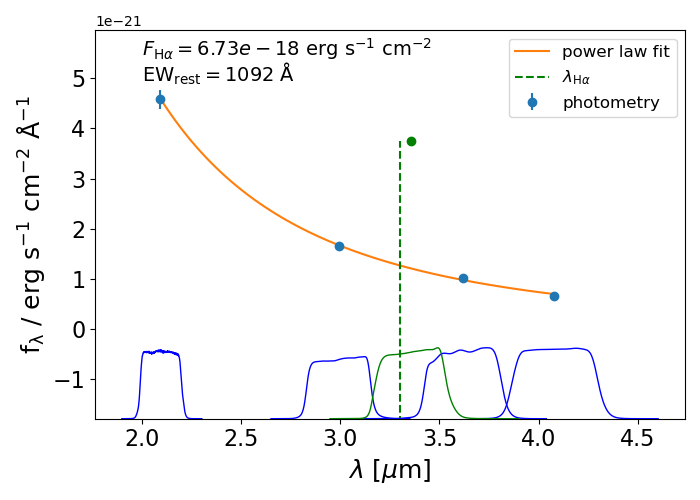}
\end{minipage}


\begin{minipage}[t]{0.68\textwidth}
    \centering
    \includegraphics[width=\linewidth]{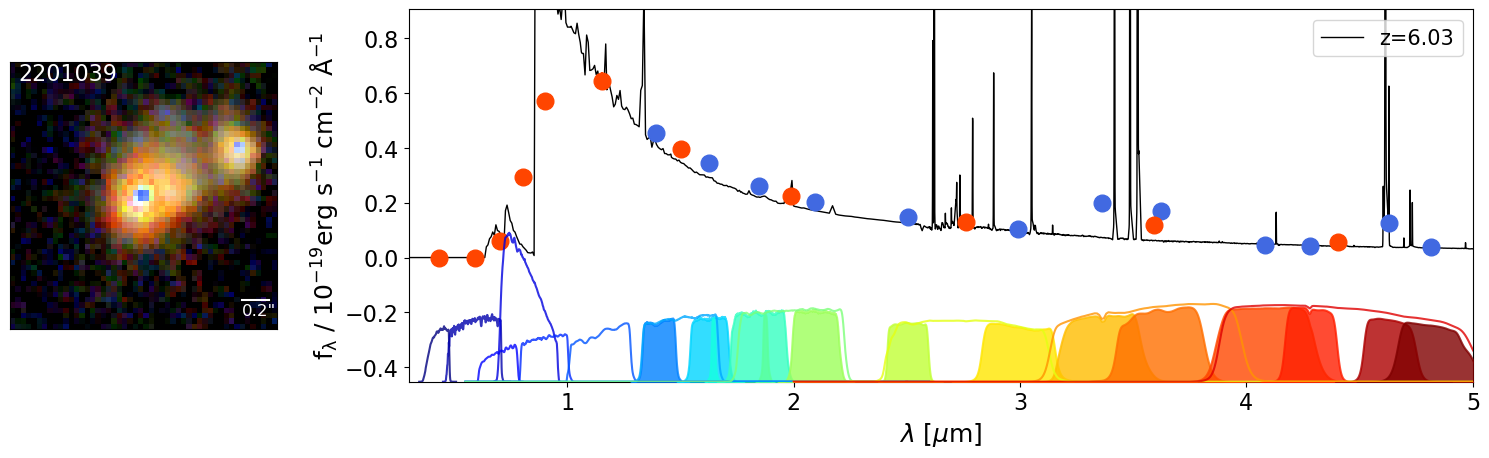}
\end{minipage}
\hfill
\begin{minipage}[t]{0.31\textwidth}
    \centering
    \includegraphics[width=\linewidth]{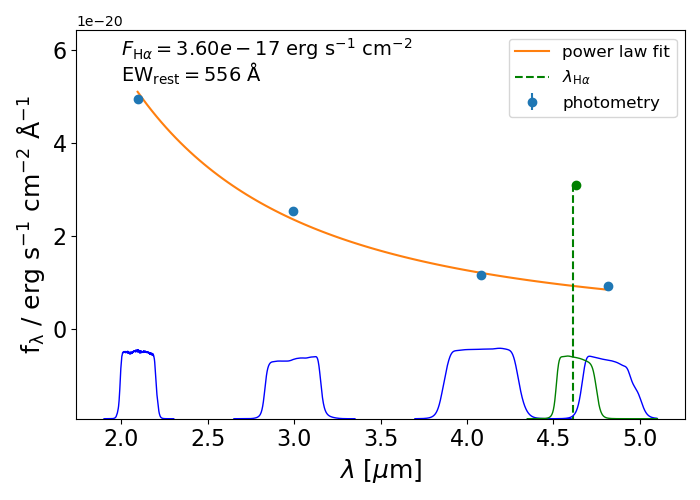}
\end{minipage}

\caption{Left: RGB (F444W, F277W, F150W) images, photometry, and best-fit EAzY models for three example \halpha-emitters detected at different redshifts. Filter response curves are shown below the spectrum, with filled curves indicating medium-bands. Photometry for wide-bands is shown in orange, for medium-bands in blue. Right: Schematic of the procedure used to measure emission line flux from photometry. The green point shows the photometry in the band containing \halpha\, while blue points show photometry in neighboring continuum bands. Photometric errors are smaller than the points. The orange curve shows a power law fit to the continuum bands. The emission line flux is taken as the flux in the \halpha\ filter after subtracting the continuum model. The wavelength of \halpha\ is shown as a vertical dashed line. The emission line flux and equivalent width are indicated.}
\label{fig:example}
\end{figure*}

\begin{figure}[ht]
\includegraphics[width=\columnwidth]{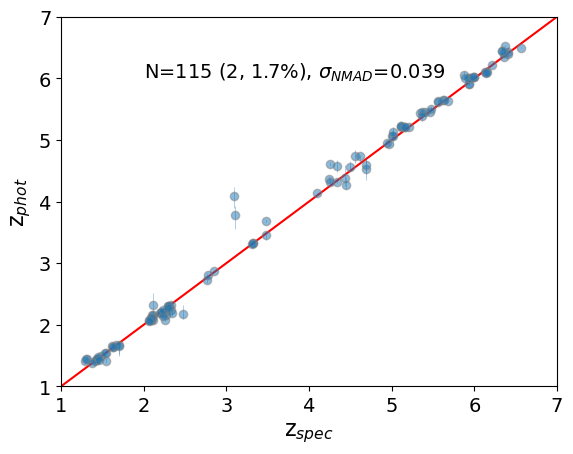}
\includegraphics[width=\columnwidth]{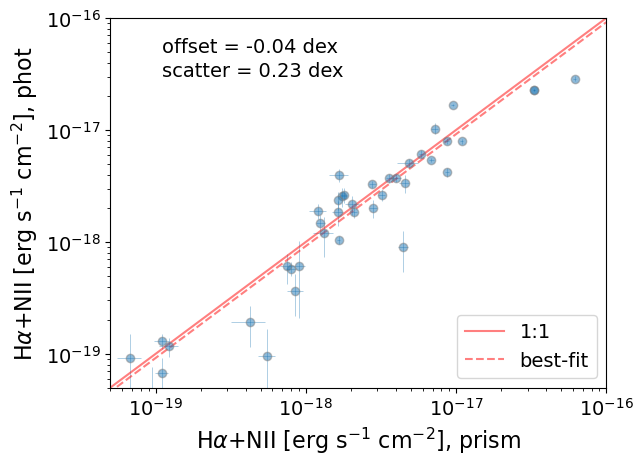}
\caption{Top: Comparison of spectroscopic redshifts (including CANUCS prism spectra and ancillary ground-based spectra, see \citetalias{sarrouh26}) with photometric redshifts. The number of sources in the comparison, outlier number and fraction, and median absolute deviation are indicated. Bottom: Comprison of \halpha+NII fluxes obtained from photometry with values derived from CANUCS prism spectra by fitting gaussian models to the lines after subtracting continuum. The offset from the 1:1 relation and scatter are indicated.}
\label{fig:spectra}
\end{figure}

\section{Data}
\label{sec:data}

We utilize the extensive multi-band imaging obtained by the CANUCS \citep[Program ID 1208, PI Willott][]{willott22} as the base of our sample. CANUCS observed five strong lensing clusters (Abell 370, MACS J0416.1-2403, MACS J1149.5+2223, MACS J1423.8+2404, and MACS J0417.5-1154) at intermediate redshift with NIRCam, NIRISS, and NIRSpec. The NIRCam flanking field observations include wide-band photometry in F090W, F115W, F150W, F277W, and F444W, and medium-band photometry in F140M, F162M, F182M, F210M, F250M, F300M, F335M, F360M, and F410M. The one exception is MACS J1149.5+2223 which lacks F162M and F250M due to a program definition error. These are supplemented by the JWST in Technicolor survey (Program ID 3362, PI Muzzin) which covers the flanking fields of the three Hubble Frontier Fields clusters (Abell 370, MACS J0416.1-2403, and MACS J1149.5+2223) with the remaining NIRCam wide- and medium-band filters (F070W, F200W, F356W, F430M, F460M, and F480M). This results in homogeneous imaging in \textit{every} wide- and medium-band NIRCam filter along with two narrow-band filters (F164N and F187N) and ancillary {\tt HST} ACS imaging from the \textit{Hubble} Frontier Fields program \citep{lotz17}. {\tt HST}/WFC3 UVIS imaging in F438W  and F606W (HST-GO16667; PI: Bradač) covers the  MACS J1423.8+2404, and MACS J0417.5-1154 flanking fields. The $3\sigma$ limiting flux densities in a $0".3$ aperture range from $\sim 2-4$ nJy in wide filters to $\sim 4-8$ nJy in medium filters ($29.5-30$ mag). Complete details of depth and completeness are provided by \citetalias{sarrouh26}. The CANUCS/Technicolor image mosaics and photometric catalogs are publicly available\footnote{https://archive.stsci.edu/hlsp/canucs}, with image reduction and photometry procedures, details on PSF measurement and homogenization, and full technical information for the surveys described in \citetalias{sarrouh26}. 

Finally, we include additional medium-band observations of the cluster pointings obtained by \jwst\ Ultimate Photometric Survey (JUMPS, PID:5890, PI:Withers). This program adds F360M, F430M, F460M, and F480M medium-band imaging to the three Frontier Fields CANUCS clusters supplementing the existing F090W, F115W, F150W, F200W, F277W, F356W, and F444W imaging. Depths are comparable to CANUCS/Technicolor. Due to the complicating effect of gravitational lensing on survey completeness, we utilize only the off-cluster NIRCam module (module A) for JUMPS (i.e. half of the observed area), where lensing effects are minimal and easily accounted for. Magnification factors reach $\mu\approx2$ in Abell 370 but are mostly $<1.5$ for the others. Image reduction and photometry are performed using the same pipeline procedures as for CANUCS/Technicolor.

The data reduction process for CANUCS/Technicolor is fully described in \citetalias{sarrouh26}. The JUMPS data are processed in the same way. Briefly, we begin the data reduction process with a modified version of the Detector1 Pipeline (calwebb\textunderscore detector1) stage of the official STScI pipeline. Astrometric alignment of the different exposures of JWST/NIRCam to HST/ACS images, sky subtraction, and drizzling to a common pixel scale of 0.04$''$ utilize the grism redshift and line analysis software for space-based slit-less spectroscopy (\texttt{Grizli}; \citealt{brammer21}). PSFs are extracted empirically by median stacking bright, isolated, non-saturated stars. Object detection and segmentation are performed on the $\chi_{mean}$ detection image created using all available NIRCam and {\tt HST} ACS images. Aperture photometry is performed on images convolved to the F444W resolution with the \texttt{Photutils} package \citep{bradley22}. We utilize 0".7 aperture photometry scaled to total based on the total to aperture ratio in the F277W filter.

\section{Analysis}
\label{sec:analysis}
\subsection{Sample Selection} 
We define 11 bins in redshift which correspond to the redshift ranges for which \halpha\ falls in the F162M, F182M, F210M, F250M, F300M, F335M, F360M, F410M, F430M, F460M, and F480M filters (see Table \ref{tab:filterbins}) at a wavelength with at least 50\% of the maximum filter transmission. Photometric redshifts  are calculated using \texttt{EAzY} \citep{brammer08}. When running EAzY we use modified templates based on the \texttt{binc100z001age6\_cloudy\_LyaReduced} template from \citet{larson23} to include stronger [OIII] emission lines to better match recent observations and the empirically-calibrated intergalactic medium/ circumgalactic medium attenuation curve of \citet{asada25} which reduces redshift bias for Epoch of Reionization galaxies. We limit our sample to sources detected at a significance of at least $5\sigma$ in the line filter and $3\sigma$ in the continuum filters (see below). We also apply a redshift quality cut by requiring a small uncertainty on the photometric redshift ($ (z_{phot,84}-z_{phot,16}) \leq 0.5$) to ensure that the \halpha\ line falls within the appropriate filter. The top panel of Figure \ref{fig:spectra} shows the comparison of photometric and spectroscopic redshifts for our sample. While the spectroscopic sample is biased toward brighter sources, the excellent correspondence and small outlier fraction ($<2\%$) suggest our sample is highly robust against interlopers. 

Table \ref{tab:ha_schechter_obs} shows the redshift range and number of sources corresponding to each filter that enter the final sample to be used in the calculation of the luminosity function. The full sample contains 4101 sources from $1.35<z<6.57$ covering a luminosity range $\rm log(L_{H\alpha} [erg\ s^{-1}]) \sim 40.7-43$ after applying cuts in completeness (see Section \ref{sec:completeness}). For the observed luminosity function, we do not remove active galactic nuclei (AGN) so that we preserve the full luminosity density. When we compute the SFR function in Section \ref{sec:SFRF}, we remove ``little red dots'' using the selections of \citet{kokorev24a, kocevski24}. During visual inspection of the highest luminosity sources, we also remove a small number of likely AGN (20) at $z<2$ based on their steep rising IR continuum past the $1.6\ \mu$m stellar bump. We check the remaining sample against the Chandra Source catalog \citep{evans24} for potential AGN and exclude two additional sources. 


 


\subsection{Emission Line Flux Measurement}

 \begin{figure*}[ht]
\centering
\includegraphics[width=\textwidth]{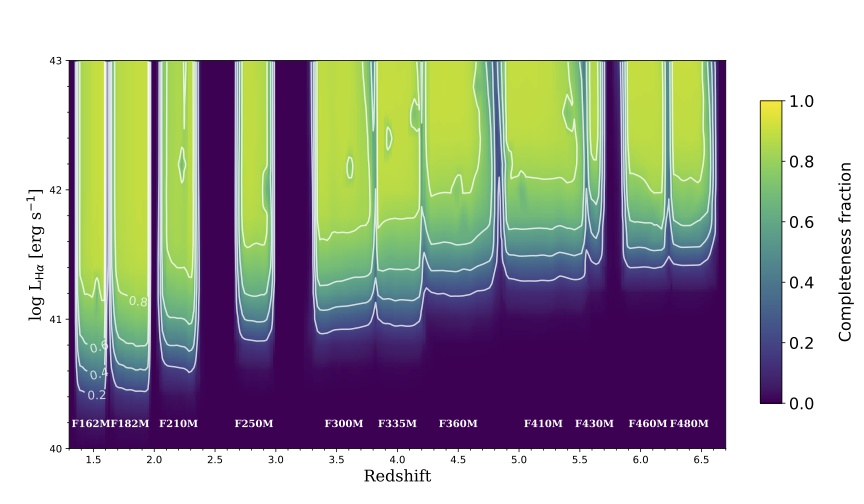}
\caption{Completeness as a function of \halpha\ luminosity and redshift for each NIRCam medium-band sample.}
\label{fig:completeness}
\end{figure*}

To measure line fluxes, we employ similar methods as those employed in other medium-band studies of emission line galaxies (ELGs) \citep[e.g.,][]{withers23, lorenz25}. For each redshift bin, we select five filters, one central filter which contains the \halpha+[NII] emission line complex, and four continuum filters which are required to be free of strong emission lines (i.e., [OIII]+H$\beta$ in the blue filters). When possible, we select two continuum filters blue-ward of the line, and two red-ward of the line. Otherwise (e.g. when \halpha\ falls in  our bluest or reddest filters), we select the four continuum filters nearest in wavelength to the line filter. The continuum level at the location of \halpha\ is measured by fitting a power law to the continuum filters in $f_\lambda$. The excess over this continuum level in the observed flux density in the filter containing the line is considered to arise from the emission line complex. The emission line flux is obtained by multiplying the excess flux density by the width in the emission line filter. The equivalent width (EW) is then given by the ratio of the emission line flux to the continuum flux density. Errors are calculated with Monte Carlo simulations. The continuum photometric flux densities are randomly perturbed by their uncertainties 100 times and the line fluxes and EWs are recalculated. The $1\sigma$ uncertainties are taken to be the 16th and 84th percentiles of these draws. Here we use only the line fluxes, an expanded analysis incorporating analysis of EWs will be presented by Martis et al. in prep. A schematic of the measurement is shown in the right panel of Figure \ref{fig:example}.

To assess the validity of our photometric redshift and emission line flux measurements, we compare with values measured from NIRSpec prism spectroscopy for all available sources in the CANUCS sample. Figure \ref{fig:spectra} shows the comparison of redshifts and line fluxes between our photometric measurements and those obtained by spectral fitting. Spectroscopic redshift measurements for NIRSpec prism observations are performed with \texttt{msaexp} \citep{brammer22}, and ancilliary ground-based redshifts are taken from the compilation in \citetalias{sarrouh26}. We find excellent agreement between photometric and spectroscopic redshifts for our sample \citepalias[see][for a comparison with the full CANUCS sample]{sarrouh26}, with an outlier fraction of less than $2\%$. For the flux comparison, we use only the NIRSpec prism spectroscopic subsample. The spectra are corrected for slit losses by scaling the spectrum to the observed photometry. We measure the line fluxes by fitting a three gaussian model to the \halpha\ and [NII] doublet emission line complex since the features are blended in the low-resolution prism spectra. We find good agreement between the spectroscopic and photometric line fluxes, with a systematic offset of only $\sim 0.04$ dex toward lower fluxes and a scatter of $\sim 0.2$ dex. Finally, we note that the fit of every source with log $L_{H\alpha}>42.5$ L$_\odot$ was visually inspected to ensure minimal contribution of interlopers at the bright end which is more sensitive to such cases.

Since \halpha\ is blended with [NII] in both the photometric and prism spectroscopic line flux measurements, we must remove the [NII] contribution to retrieve isolated \halpha\ fluxes. We determine the contribution of [NII] through the use of the redshift-dependent mass-metallicity relation measured by \citet{Isobe26}. For this calculation, we use the stellar masses derived from \texttt{DENSEBASIS} \citep{iyer19} that are presented in \citet{sarrouh26}. The metallicity is calculated for each galaxy using the mass-metallicity relation and used to correct for the contribution of [NII]. For our two lowest redshift bins, this correction can be substantial (up to $\sim 20\%$), but for the rest of the sample at both higher redshift and lower mass, metallicity is in general low, so the removed contribution of [NII] constitutes only $\sim 5\%$ of the total flux of the emission line complex. 

Due to their on-sky proximity to the foreground lensing clusters, our sample undergoes minor, but non-negligible flux magnification (typically less than a factor of two in the flanking fields and off-cluster modules utilized here). All fluxes and luminosities reported here are corrected for magnification using the lens models described in the CANUCS DR1. This small degree of magnification also produces a small effect on the surveyed volume. We account for this reduced volume in our completeness simulations (see Section \ref{sec:completeness}).

\begin{figure*}[ht]
\centering
\includegraphics[width=\textwidth]{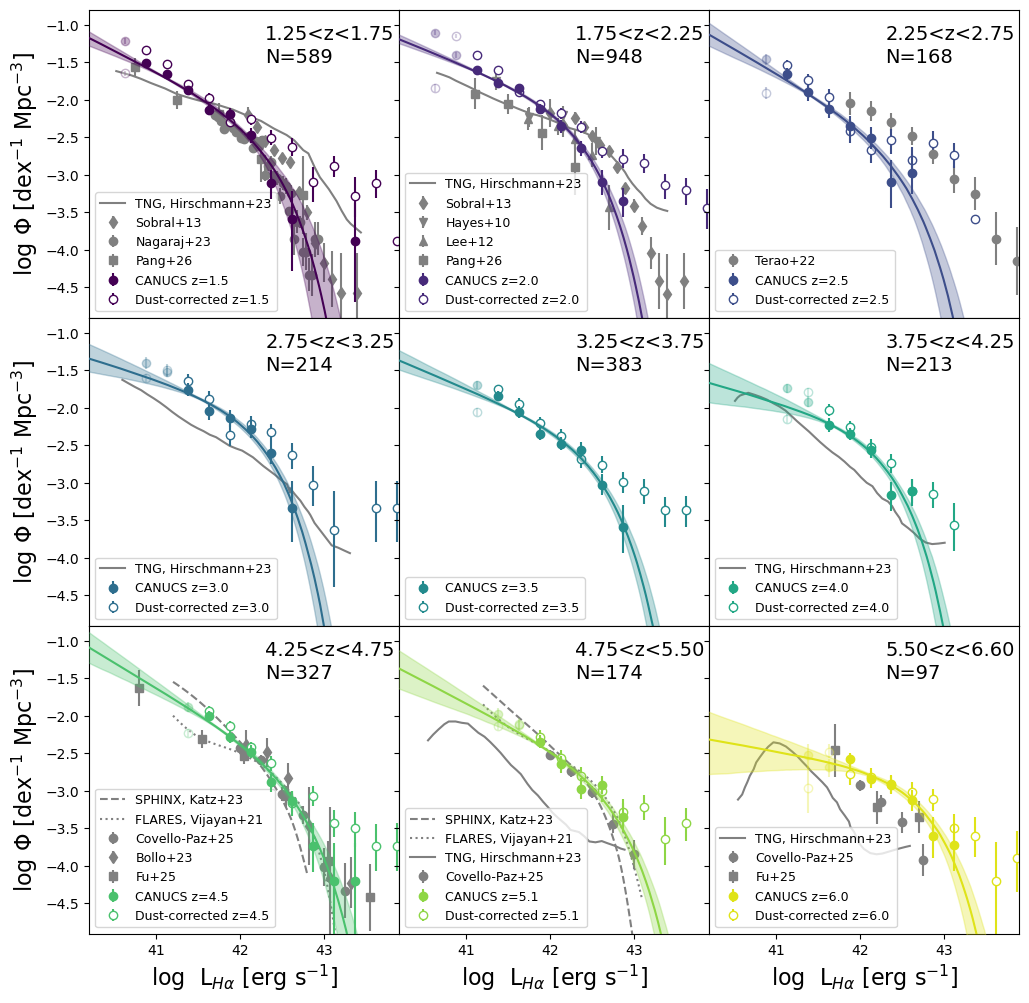}
\caption{\halpha\ luminosity function (colored points) and Schechter fits (colored curves) in each of our nine redshift bins. Open circles show the dust-corrected measurements. Transparent circles of both types indicate bins below the 50\% completeness limit and are excluded from the Schechter fitting. Previous observations including ground-based surveys \citep{sobral13, hayes10, lee12, terao22}, \textit{Spitzer} \citep{bollo23}, \hst grism \citep{nagaraj23},  \jwst\ NIRISS \citep{pang26} and  JWST NIRCam grism surveys \citep{covelo-paz25, fu25} are shown as grey symbols. We also show predictions from recent cosmological simulations including Illustris TNG \citep{hirschmann23}, FLARES \citep{lovell21, vijayan21}, and SPHINX \citep{katz23} as gray curves. The redshift range and number of sources contributing to the luminosity function Schechter fit in each bin are indicated in the top of each panel.}
\label{fig:LF}
\end{figure*}

\subsection{Completeness Calculations}
\label{sec:completeness}
To accurately measure any luminosity function, one must properly account for the presence of sources which exhibit brightness in the range of interest, yet fail to meet the selection criteria, i.e., the survey completeness. Since we combine multiple surveys with somewhat different depths and filter sets, the completeness of our photometric-redshift-selected sample is not trivially determined. Indeed, the question of whether any given galaxy is detected by an image detection pipeline is more straightforward, and more complicated completeness effects are introduced by the quality cuts used in the sample selection. To ensure proper treatment of completeness, we perform full simulations of artificial sources inserted into our images and testing their recoverability by our sample selection. The procedure is based off that presented in \citet{willott24} to calculate completeness corrections to the UV luminosity function. 

We create images of artificial galaxies using S\'ersic profiles. The profile parameters are sampled from Gaussian distributions of ellipticities centered at 0.3 with $\sigma=0.2$ (values below 0 and above 0.7 excluded) and S\'ersic indices centered at 1.5 with $\sigma=0.3$ (values below 0 excluded). The sizes are assigned using the size-luminosity-redshift relation from \citet{morishita24} which is derived from a compilation of Cycle 1 JWST data. 

SEDs for the mock galaxies are generated using \texttt{BAGPIPES} using a wide range of stellar population parameters. The synthetic galaxies utilize a delayed star formation history (SFH) sampling over age, $\tau$ (ensuring both rising and falling SFHs), stellar mass, ionization parameter, and $V$-band attenuation, \Av. We construct a total of 240 template SEDs and verify that they cover the full range of our observed line fluxes, equivalent widths, and flux densities in the selection bands. After synthetic sources are inserted into the images, we run the same photometric pipeline as our original data processing to generate detection images and photometric catalogs including the fake sources. We re-run our source selection using the photometry of the synthetic sources and check the ratios of recovered 0.3$''$ aperture fluxes to simulated total fluxes are within a sensible range (0.1 to 1.4). The recovered fraction of sources as a function of input redshift and \halpha\ luminosity then determines the completeness level. 

A final adjustment to account for the reduced volume due to magnification is performed as follows. We use the CANUCS lensing maps to calculate magnification for simulated sources that were successfully detected and selected. This distribution gives the area per magnification bin. The completeness array is shifted in luminosity in proportion to the effective area to de-lens the luminosities.

Figure \ref{fig:completeness} shows the completeness as a function of redshift and \halpha\ luminosity for 11 redshift bins determined by each medium-band filter. We find that our dataset reaches 50\% completeness at log(L/L$_\odot$) = 40.7 in our lowest redshift bin, and 41.6 in the highest. This is more than an order of magnitude deeper in \halpha\ luminosity than previous ground-based measurements at low redshift, and comparable to recent NIRCam grism measurements at high redshift with a similar exposure time  ($\sim2$ hrs/pointing).

\section{Results}
\label{sec:results}
\subsection{Observed \halpha\ Luminosity Function at $1.3<z<6.6$}
\label{sec:observed}
We compute the \halpha\ luminosity function over our full redshift range using two different redshift binning schemes. For ease of comparison with simulations and observations with instruments other than \jwst\ NIRCAM, we here show nine bins with redshift widths of 0.5 at $z<5$ and 1.0 at $z>5$. When the redshift range corresponding to a given medium-band filter is not fully contained in a single redshift bin, we use only the overlapping redshift range. Thus the \halpha-emitter samples for some filters are spread across multiple redshift bins. In Appendix \ref{sec:appendix} we show the luminosity function directly measured in 11 redshift bins corresponding to the redshift ranges covered by each filter. We use the $1/V_{max}$ formalism of \citet{schmidt68}:
\begin{equation}
    \Phi(L) = \frac{1}{d\ \rm log\ L} \sum_i \frac{V}{V_{max,i}C_i}
\end{equation}
where L is \halpha\ luminosity, $V$ is the full survey volume in a given redshift bin, $V_{max},i$ is the maximum volume in which source $i$ could be detected, and $C_i$ is the completeness for source $i$ according to its redshift and luminosity. We note that the observable volume does not always include the full redshift range of the bin due to the wavelength coverage of the medium-bands and this effect is fully accounted for. We calculate errors by bootstrapping the sample within each redshift bin 1000 times while simultaneously perturbing the \halpha\ luminosity by its uncertainty. This accounts for sample variance and naturally accounts for Eddington bias since faint sources will be preferentially scattered into higher luminosity bins by perturbing the fluxes. We implement an error floor for bins with fewer than five sources using the corrections to poisson statistics of \citet{gehrels86}. Additionally we estimate errors from cosmic variance by using the online calculator of \citet{trenti08} and add them in quadrature. We note that since we sample 3-5 distinct fields depending on filter coverage, cosmic variance is a minor formal contribution to the uncertainty (but see Section \ref{sec:cv}). 

We also compute the dust-corrected luminosity functions as follows. For each source, we utilize the stellar \Av\ determined from SED modeling with \texttt{DENSEBASIS} (see \citealt{sarrouh26} for details) which assumes a \citet{calzetti00} dust attenuation curve. The intrinsic \halpha\ luminosity is then given by 
\begin{equation}
    L_{H_\alpha,int} = L_{H_\alpha,obs} \times 10^{0.4A_{H\alpha}},
\end{equation}
where A$_{H\alpha}$ is the nebular attenuation at the wavelength of \halpha. We assume a stellar to nebular attenuation ratio of 0.44. Since the completeness of the sample is based on observed \halpha\ luminosity, we use the identical sample to calculate the dust-corrected luminosity function. 

Figure \ref{fig:LF} shows the completeness-corrected \halpha\ luminosity function in nine redshift bins as colored points. Open symbols designate the dust-corrected values. For both cases, we plot luminosity bins where completeness is less than $50\%$ as semi-transparent points. The best-fit Schechter functions to the observed luminosity functions (see Section \ref{sec:schechter}) are shown as curves of corresponding color. Other measurements of the \halpha\ luminosity function from ground-based surveys at $z \sim 1-3$ \citep{sobral13, hayes10, lee12, terao22}, \textit{Spitzer} \citep{bollo23}, \hst\ grism \citep{nagaraj23},  \jwst\ NIRISS \citep{pang26} and  JWST NIRCam grism surveys \citep{covelo-paz25, fu25} are shown as grey symbols. We also show predictions from recent cosmological simulations including Illustris TNG \citep{hirschmann23}, FLARES \citep{lovell21, vijayan21}, and SPHINX \citep{katz23}.  

We find good agreement with previous measurements in most redshift bins. The largest discrepancy occurs in our $z \sim 2.5$ luminosity function, in which we find a $0.2-0.5$ dex lower normalization in the overlapping luminosity bins compared to \citet{terao22}, though we note their measurement is also significantly higher than the \citet{sobral13} measurement at $L_{H\alpha}>10^{43}\ \rm erg\ s^{-1}$ at only slightly lower redshift. We also find a slightly higher normalization in the $z \sim 6$ bin than \citet{covelo-paz25, fu25}, though within errors. At $z>3$ we find higher normalization than the ILLUSTRIS TNG simulation in addition to steeper slopes at lower redshift. In contrast, the agreement with the SPHINX and FLARES predictions at $4<z<6$ is much better. Our luminosity functions showcase the power of \jwst\ medium-bands to probe the luminosity function over a wide range of redshifts and luminosities, competing effectively with \jwst\ NIRCam grism measurements.

\begin{figure}[ht]
\includegraphics[width=\columnwidth]{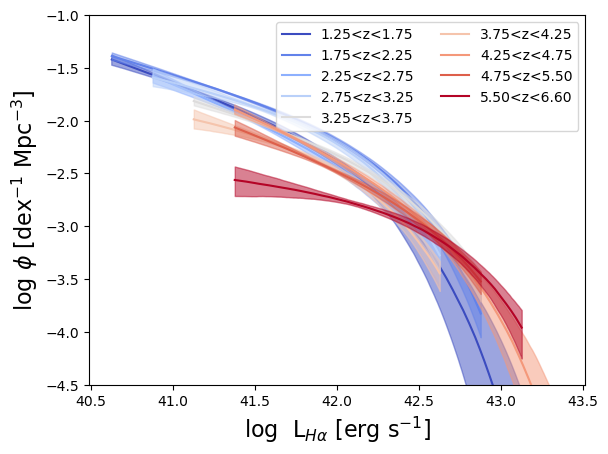}
\includegraphics[width=\columnwidth]{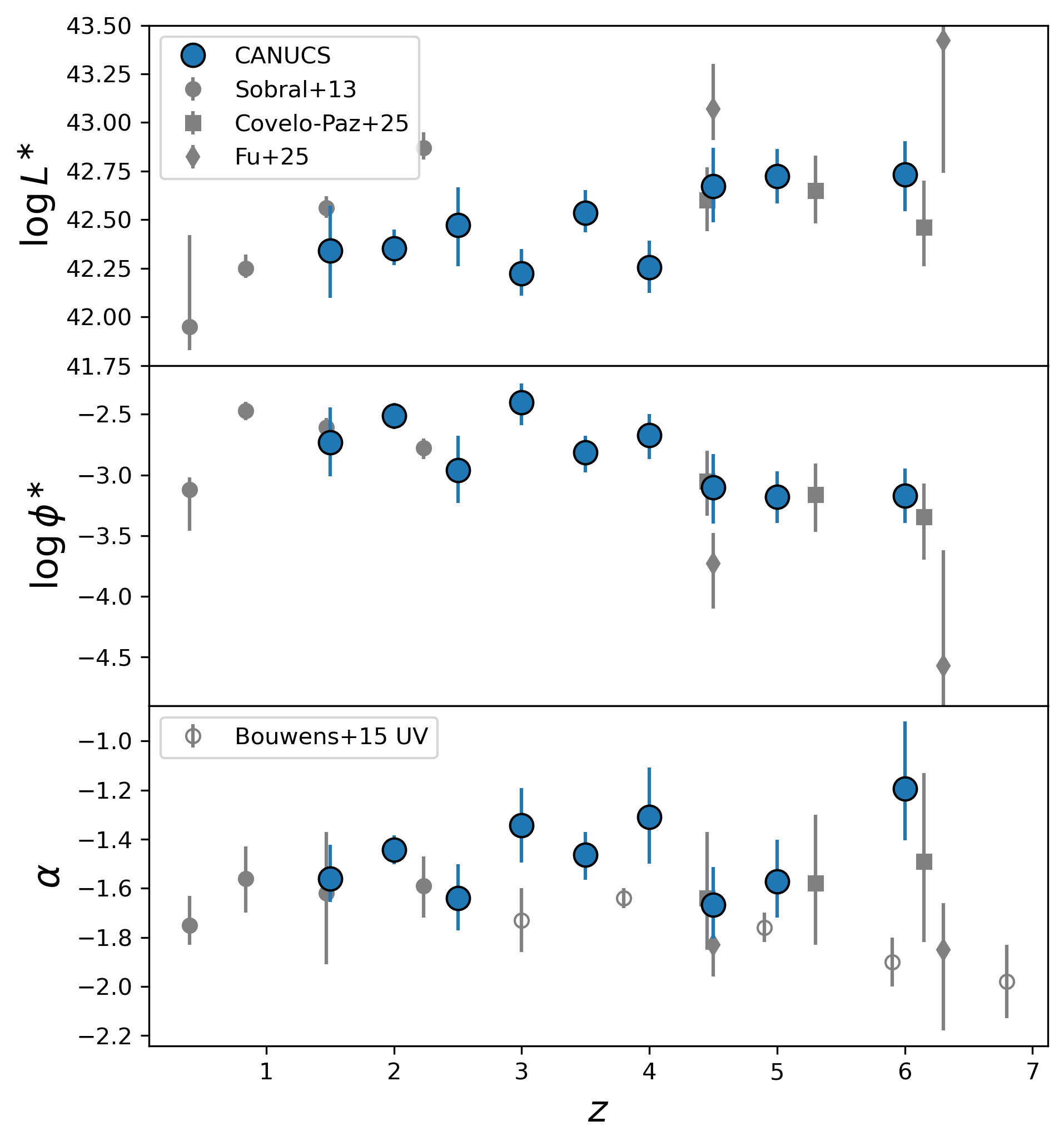}
\caption{Top: Schechter fits to the \halpha\ luminosity function in every redshift bin shown together in order to better illustrate the evolution. Bottom: Redshift evolution of the best-fit Schechter function parameters for the observed \halpha\ luminosity function. Uncertainties are determined through 1000 Monte Carlo simulations remeasuring the fit to the luminosity after perturbing emission line fluxes by their uncertainties and bootstrapping the sample. Uncertainties from cosmic variance are added in quadrature. Measurements from \citet{sobral13, covelo-paz25, fu25} are shown for comparison. We also show the evolution of the faint-end slope of the UV luminosity function from \citet{bouwens15}.}
\label{fig:LF-z}
\end{figure}

\subsection{Redshift Evolution}
\label{sec:schechter}

\begin{table*}
\centering
\begin{tabular}{lccrccc}
\toprule
$z_{mid}$ & $z_{range}$ & $Filters$ & $N$ & $\log L^*$ & $\log \phi^*$ & $\alpha$ \\
\midrule
1.50 & $1.25 < z < 1.75$ & F162M, F182M & 761 & $42.34^{+0.23}_{-0.24}$ & $-2.73^{+0.29}_{-0.28}$ & $-1.56^{+0.14}_{-0.10}$ \\
2.00 & $1.75 < z < 2.25$ & F182M, F210M & 1316 & $42.35^{+0.09}_{-0.09}$ & $-2.51^{+0.11}_{-0.11}$ & $-1.44^{+0.06}_{-0.06}$ \\
2.50 & $2.25 < z < 2.75$ & F210M, F250M & 229 & $42.47^{+0.19}_{-0.21}$ & $-2.96^{+0.28}_{-0.27}$ & $-1.64^{+0.14}_{-0.13}$ \\
3.00 & $2.75 < z < 3.25$ & F250M & 293 & $42.22^{+0.13}_{-0.12}$ & $-2.40^{+0.15}_{-0.19}$ & $-1.34^{+0.15}_{-0.15}$ \\
3.50 & $3.25 < z < 3.75$ & F300M & 462 & $42.53^{+0.12}_{-0.10}$ & $-2.82^{+0.14}_{-0.17}$ & $-1.46^{+0.09}_{-0.10}$ \\
4.00 & $3.75 < z < 4.25$ & F300M, F335M, F360M & 331 & $42.25^{+0.14}_{-0.13}$ & $-2.67^{+0.17}_{-0.20}$ & $-1.31^{+0.20}_{-0.19}$ \\
4.50 & $4.25 < z < 4.75$ & F360M & 372 & $42.67^{+0.20}_{-0.19}$ & $-3.10^{+0.28}_{-0.29}$ & $-1.67^{+0.15}_{-0.14}$ \\
5.00 & $4.75 < z < 5.50$ & F360M, F410M, F430M & 219 & $42.72^{+0.14}_{-0.14}$ & $-3.18^{+0.21}_{-0.22}$ & $-1.57^{+0.17}_{-0.15}$ \\
6.05 & $5.50 < z < 6.60$ & F430M, F460M, F480M & 118 & $42.73^{+0.17}_{-0.19}$ & $-3.17^{+0.23}_{-0.23}$ & $-1.19^{+0.27}_{-0.21}$ \\
\bottomrule
\end{tabular}
\caption{Schechter parameters for the observed H$\alpha$ luminosity function measured in each redshift bin. $z_{mid}$ is the median redshift of the bin. $N$ is the number of sources included in the Schechter fit, which uses only sources with luminosity above the $>50\%$ completeness level (excludes sources in the transparent bins in Figure \ref{fig:LF-z}.}
\label{tab:ha_schechter_obs}
\end{table*}

We also perform an unbinned extended maximum-likelihood fit \citep{sandage79, marshall83} to fit a Schechter \citep{schechter76} functional form to our data in each redshift bin. The Schechter function, characterized by the characteristic turnover luminosity $L^*$, normalization $\Phi^*$, and faint-end slope $\alpha$, is defined as follows
\begin{equation}
 \Phi(L)dL = \Phi^*\left(\frac{L}{L^*}\right)^\alpha e^{-L/L*} \frac{dL}{L^*}
\end{equation}
We use Markov Chain Monte Carlo fitting to determine the best-fit parameters and uncertainties, performing 1000 iterations. As with the binned luminosity function, we perturb \halpha\ luminosities by their errors and bootstrap the sample in each realization to determine the errors. Uncertainties from cosmic variance are incorporated directly in the Monte Carlo by perturbing each realization with a field-combined fractional cosmic variance term estimated from the \citet{trenti08} calculator, combining the CANUCS pointings as independent volumes following \citet{moster11}. 

The top panel of Figure \ref{fig:LF-z} shows the fitted Schechter functions together in order to better illustrate the evolution. The bottom panel shows the evolution of $\phi^*$, $L^*$, and $\alpha$ with redshift. We observe a peak in the normalization $\phi^*$ at $z\sim3$, visible in both the plotted Schechter functions and the $\phi^*$ parameter, followed by a decrease toward higher redshift. We find tentative evidence for an increase in the characteristic luminosity $L^*$, with redshift, though the degeneracy with $\phi^*$ makes this uncertain. We also note on this point that our $L^*$ measurements fall below those of \citet{sobral13} which extend to brighter luminosity at the low redshift end. 
Finally, we observe no strong trend of $\alpha$ with redshift, with typical slopes ranging from -1.2 to -1.6 . This is in good agreement the slopes of $\sim -1.6$ at lower redshift by \citet{sobral13} and  \text{-1.49 -- -1.65} found by \citet{covelo-paz25} at $4<z<6$. Though \citet{fu25} find slightly steeper slopes of -1.85 at $4<z<6$, being able to measure down to very faint luminosities thanks to gravitational lensing. This supports earlier findings that the faint-end slope of the \halpha\ luminosity function does not significantly evolve with redshift. Moreover, along with \citet{pang26}, we demonstrate the slope stays constant with no turnover all the way down to $L_{H\alpha}=10^{40.5}\ \rm erg\ s^{-1}$ at $z<2$, more than an order of magnitude below previous ground-based determinations. Interestingly we find slightly shallower faint-end slopes when comparing to the UV luminosity function of \citet{bouwens15}. The shallower \halpha\ slopes might suggest lower detectability at low star SFRs due to sensitivity to star formation on a shorter timescale, or potentially differences between stellar and nebular attenuation in low-mass, high-z galaxies. The fitted parameters for each of the Schechter functions are also shown in Table \ref{tab:ha_schechter_obs}.  

\subsection{Star Formation Rate Functions}
\label{sec:SFRF}

\begin{figure*}[ht]
\centering
\includegraphics[width=\textwidth]{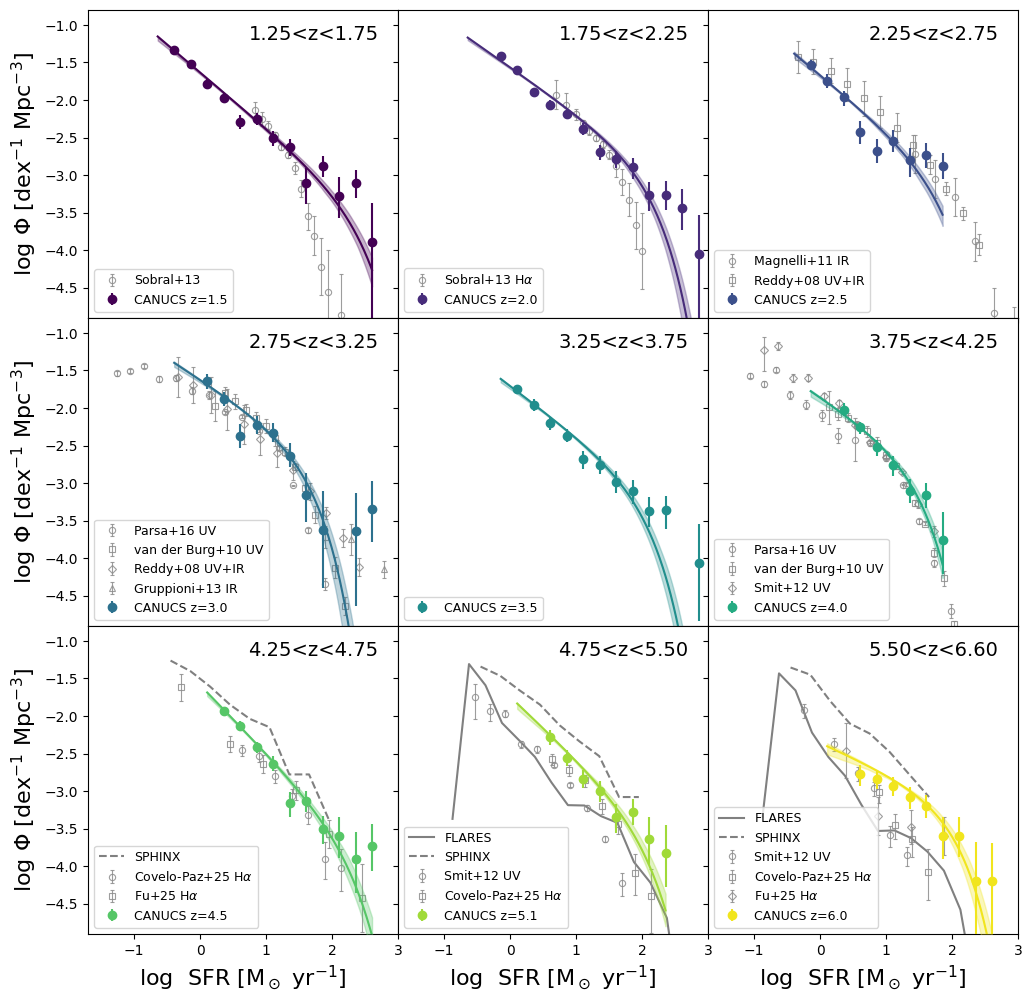}
\caption{Star formation rate functions and Schechter fits in the same redshift bins as the luminosity functions. For comparison we show measurements from \halpha\ \citep{sobral13, covelo-paz25, fu25}, a compilation of literature results from UV and IR \citep{reddy08, magnelli11, smit12, gruppioni13, parsa16}, and predictions from the FLARES \citep{lovell21, vijayan21} and SPHINX \citep{katz23} simulations.} 
\label{fig:SFRF}
\end{figure*}

\begin{figure}[ht]
\includegraphics[width=\columnwidth]{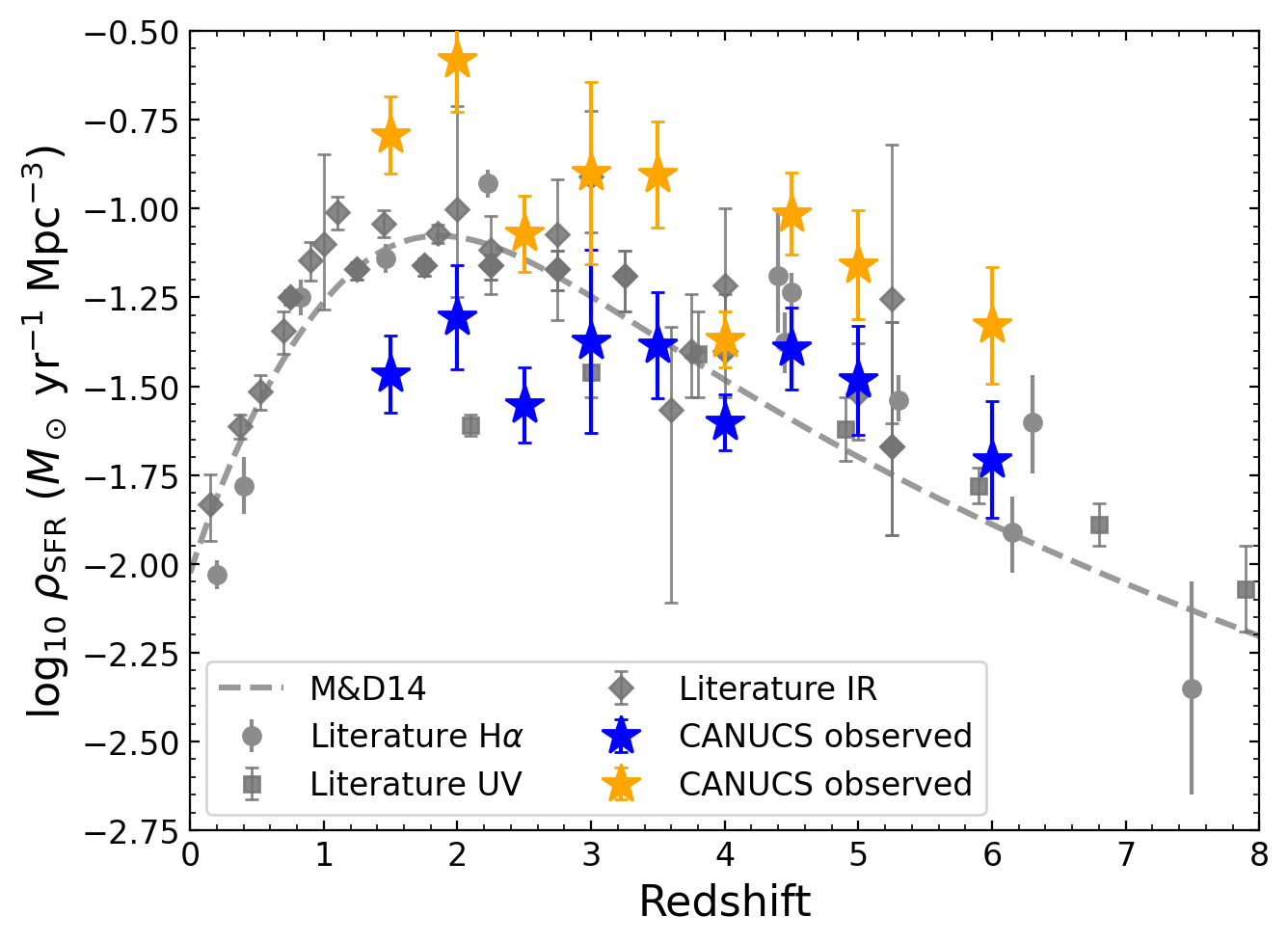}
\caption{Star formation rate density as a function of redshift obtained from integrating the star formation rate functions to a limit of $0.27\ \rm M_\odot\ yr^{-1}$.
The observed SFRD is shown as blue stars and the dust-corrected SFRD as orange stars. For comparison we show a compilation from the literature with various sources including \halpha\ \citep{sobral13, rinaldi24, covelo-paz25, fu25}, UV \citep{bouwens22UVLF} and IR \citep{gruppioni13, gruppioni20, traina24, sun25}. The reference relation of \citet{madau14} is plotted as a dashed curve.} 
\label{fig:SFRD}
\end{figure}

We convert our-dust corrected luminosity measurements to measure SFR functions in the same redshift bins. We note that the (observed and corrected) luminosity functions contain AGN since we wish to present the \halpha\ from all sources. Before computing the SFR functions we remove several classes of AGN including little red dots using the selections of \citet{kokorev24a, kocevski24}, x-ray sources from the Chandra point source catalog \citep{evans24}, and obscured AGN identified during our visual inspection of bright sources. In addition to all sources with observed $L_{H\alpha}>10^{42.5}\ \rm erg\ s^{-1}$, we also inspected all any sources which contributed more than 10\% of the integrated SFRD in any redshift bin. We removed 20 sources exhibiting very red optical slopes in conjunction with point-like morphology at long wavelength, excluding clearly AGN-dominated systems from our sample. In all, 104 sources are removed by the AGN/LRD cuts. Due to the lack of rest-frame mid-IR coverage, a more systematic removal of obscured AGN is unfortunately not feasible. Medium-band surveys with overlapping MIRI coverage will allow a more complete characterization of the contribution of AGN to the \halpha\ luminosity function. We defer this investigation to Martis et al. in prep. which will report the \halpha\ luminosity function from the Medium-band Imaging with NIRCam to Explore ReVolutionary Astrophysics (MINERVA, \citealt{muzzin25}) survey. 

We adopt the conversion between \halpha\ luminosity and SFR from \citet{kennicutt12, hao11, murphy11} such that
\begin{equation}
    \rm log(SFR[M_\odot\ yr^{-1}]) = log(L_{H\alpha}[erg\ s^{-1}]) - 41.27.
\end{equation}
We fit Schechter functions to the SFR functions in a similar manner as the \halpha\ luminosity functions, fitting in dust-corrected luminosity space and converting to SFR afterwards. The one change we implement is to provide a prior on the SFR function parameters based on the observed luminosity function priors, since we found that flat priors led to poor fits due to the elevated bright end (see discussion below). This led to degeneracy in the fitted parameters, therefore we do not report the Schechter fit parameters for the SFR functions. Figure \ref{fig:SFRF} shows the SFR functions measured in the same nine redshift bins as the luminosity functions as well as those reported from conversion of the \halpha\ luminosity functions by \citet{sobral13} and \citet{covelo-paz25}. We also show a compilation of literature results from UV and IR measurements \citep{reddy08, magnelli11, smit12, gruppioni13, parsa16} and predictions from the FLARES \citep{lovell21, vijayan21} and SPHINX \citep{katz23} simulations. 

We find shallower faint-end slopes and a larger turnover SFR than \citet{sobral13} in our two lowest redshift bins. This may be partially due to our dust correction prescription, which operates on a per-object basis shifting sources from low to high luminosity and therefore flattening the faint-end slope, rather than the constant shift from a single attenuation correction used in \citet{sobral13}. We interpret this as evidence of real differences in the required dust corrections at the faint and bright ends of the luminosity function. Since our selection is based on signal-to-noise in photometry (including continuum) rather than line flux, bright sources with faint observed \halpha\ are retained in our sample. In particular, the exquisite photometry covering $0.4-5\ \mu \rm m$ allows us to confidently include heavily obscured sources, which may have weak observed \halpha\ fluxes in our sample thanks to confident photometric redshifts from the detection of other spectral features such as the Balmer break (see Figure \ref{fig:example}).  The combination of a large intrinsic \halpha\ luminosity with the small observable volume gives such sources large weights in the $1/V_{max}$ formalism, leading to large $\phi$ in high-luminosity bins. This is readily apparent when comparing our observed and corrected luminosity functions in Figure \ref{fig:LF}. Due to the relatively small survey area of CANUCS/Technicolor/JUMPS ($\sim 40-60\ \rm arcmin^2$ depending on filter), our ability to fully constrain the bright end turnover is still somewhat limited. This shortcoming will be addressed with larger medium-band surveys such as MINERVA. 

At $z>2.25$ we agree better with previous determinations of the SFR function from both \halpha\ and UV and IR. The one exception is the $5.5<z<6.6$ bin in which we find an excess of high-SFR sources, potentially driven by an overdensity at this redshift in Abell 370 (see Section \ref{sec:cv}). As with our \halpha\ luminosity functions, we also compare to predictions from SPHINX and FLARES in overlapping redshift ranges. Our faint-end slopes and normalization are similar to the SPHINX predictions in the $z=4-5$ range, though fall lower at $z=6$. On the other hand, our observations fall at $\sim0.5$ dex higher normalization than the FLARES predictions.

\section{Discussion}
\label{sec:Discussion}

\subsection{Evolution of the Star Formation Rate Density}

\begin{table}
\centering
\caption{CANUCS cosmic star formation rate density measurements from dust-corrected \halpha.}
\label{tab:sfrd}
\begin{tabular}{ccc}
\toprule
$z_{mid}$ &$z_{range}$ & $\log_{10}(\rho_{\rm SFR}/{\rm M_\odot\,yr^{-1}\,Mpc^{-3}})$ \\
\midrule
1.50 & $1.25<z<1.75$ & $-0.92_{-0.06}^{+0.04}$ \\
2.00 & $1.75<z<2.25$ & $-0.81_{-0.04}^{+0.04}$ \\
2.50 & $2.25<z<2.75$ & $-1.15_{-0.04}^{+0.04}$ \\
3.00 & $2.75<z<3.25$ & $-1.05_{-0.04}^{+0.04}$ \\
3.50 & $3.25<z<3.75$ & $-1.04_{-0.03}^{+0.04}$ \\
4.00 & $3.75<z<4.25$ & $-1.37_{-0.02}^{+0.03}$ \\
4.50 & $4.25<z<4.75$ & $-1.10_{-0.03}^{+0.03}$ \\
5.00 & $4.75<z<5.50$ & $-1.24_{-0.02}^{+0.03}$ \\
6.00 & $5.50<z<6.60$ & $-1.38_{-0.04}^{+0.06}$ \\
\bottomrule
\end{tabular}
\end{table}

We integrate the SFR functions down to a limit of $0.27\ \rm M_\odot\ yr^{-1}$ 
in order to measure the SFR density as a function of redshift consistently with previous measurements \citep{covelo-paz25, fu25}. We verified that using an integration limit based on the characteristic SFR ($0.03\ \times SFR^*$) does not significantly change our results. The fraction of the SFRD contributed by galaxies fainter than our completeness limit, and thus affected by uncertainty in the faint-end slope of the SFR functions, is $<10\%$ at $z<4$, and reaches $\sim 20\%$ at higher redshift. Figure \ref{fig:SFRD} shows our measurements (orange and blue stars) which are tabulated in Table \ref{tab:sfrd} and other results from the literature (gray symbols) compared to the \citet{madau14} fit. We have included additional jackknife variance errors to account for observed field-to-field variation beyond the \citet{trenti08} estimates (see Section \ref{sec:cv}). Literature measurements include \halpha\ measurements from \citet{sobral13,rinaldi24, covelo-paz25, fu25}, UV measurements from \citet{bouwens22UVLF}, and IR measurements from \citet{gruppioni13, gruppioni20, traina24}. Our fiducial dust-corrected measurements generally fall toward the upper envelope of the scatter of previous measurements, though we roughly reproduce the peak in SFRD at $z\sim2$ and decline toward higher redshift. The $z\sim2$ and $z\sim6$ bins falls noticeably higher than most previous literature determinations, which we investigate below. 

As a point of comparison, we also calculate the SFRD corresponding to the observed \halpha\ LF (with AGN removed). This allows us to assess the impact of our dust corrections, and to provide a more meaningful comparison with the literature. The observed SFRD fall near the UV measurements of \citet{bouwens22UVLF}, suggesting they both effectively trace the unobscured star formation component well, despite the difference in star formation history sampling. Our dust-corrected SFRD is more in line with recent IR measurements, lending confidence to our dust corrections on average despite the limitations of spectral energy distribution modeling to constrain it in individual cases without IR data. Interestingly, this exercise also enables us to estimate the obscured fraction of star formation as a function of redshift. From a maximum of 80\% obscured fraction at $z=1.5-2$, we observe a decline to $\sim 50\%$ by $z=4$, though interestingly a plateau after that point. The trend up to $z=4$ agrees well with \citet{zavala21}, though our observed plateau disagrees with the expected decline toward higher redshift. Some recent work suggests elevated obscured fractions may persist to higher redshift \citep{martis25, sun25}, but confirming this result will require more extensive IR observations of large samples, which remains difficult with current instrumentation. 

These findings suggest a potentially elevated SFRD compared to previous measurements, particularly at $z=1-2$. Interestingly, \citet{matthews24} also find evidence of a higher SFRD from recent radio measurements compared to previous UV-IR determinations at $0.2<z<1.3$. If real, this carries significant implications for the agreement of the SFRD with the measured stellar mass density, potentially requiring modifications to the assumed initial mass function, dust attenuation law, or calibrations of SFR indicators at high redshift. With the current analysis, we consider these findings tentative. The result is robust against statistical uncertainty in our dust corrections; simulations of our SFR functions with inflated dust attenuation errors confirm a $<5\%$ contribution to the SFRD due to Eddington bias. On the other hand, systematic effects such as the assumed dust attenuation law and the ratio of nebular to stellar attenuation create larger shifts (order $\sim0.2$ dex), potentially bringing our measurements more in line with previous measurements. A resolution of this issue will require detailed, multi-wavelength observations of large galaxy samples at cosmic noon to disentangle the effects of dust, star formation histories, and metallicity on our determination of stellar mass growth at this crucial epoch.

\begin{figure}[ht]
\includegraphics[width=\columnwidth]{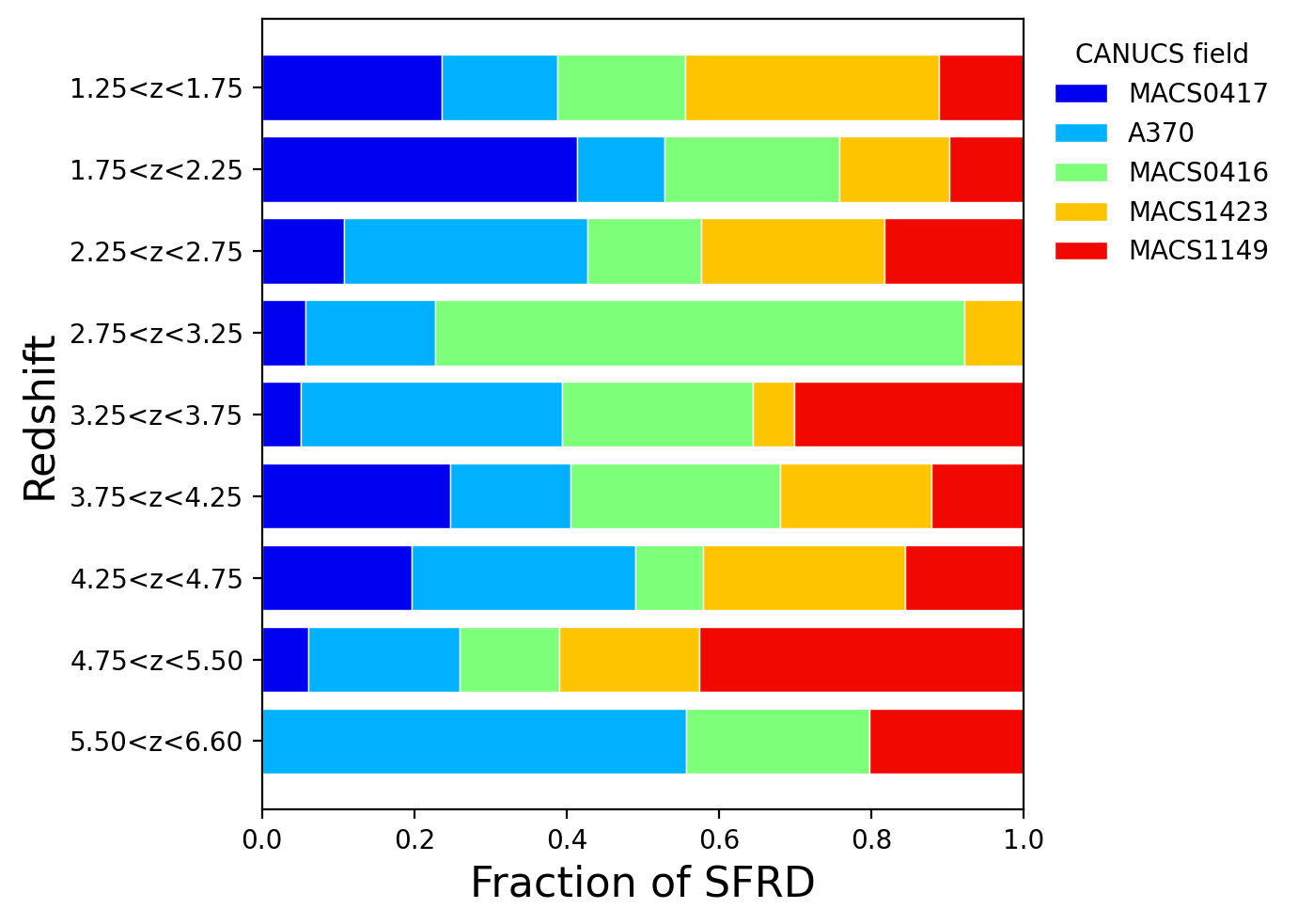}
\caption{Fractional contribution of each CANUCS field to the total measured star formation rate density. Note that observations are not available at $2.75<z<3.25$ for MACS1149 and at $5.50<z<6.60$ for MACS0417 and MACS1423.} 
\label{fig:fields}
\end{figure}

\subsection{Effects of Overdensities on the Luminosity and Star Formation Rate Functions}
\label{sec:cv}
As discussed in Section \ref{sec:SFRF}, the modest survey volume (each cluster and flanking field is a single NIRCam pointing) makes us especially sensitive to sources with large dust corrections and intrinsic \halpha\ luminosities. While visually inspecting the sources with the largest SFRD contributions, we confirm the presence a group of highly star-forming dusty galaxies in the flanking field of MACS0417 at $z=2.1$. Similarly, overdensities of \halpha-emitters in MACS1149 at $z=5.1-5.2$ \citep{wang26} and A370 at $z=6$ \citep{antwi-danso25} dominate the SFRD budget in their respective redshift bins. For the former, the low observed \halpha\ luminosity relative to the SFR (due to the large dust obscuration) grants these sources large weight in the $1/V_{max}$ formalism, resulting in the high $\Phi$ value at the bright end in the second panels of Figures \ref{fig:LF} and \ref{fig:SFRF}.

These discoveries prompted us to more closely investigate the effect of cosmic variance on our SFRD measurements. Figure \ref{fig:fields} shows the fractional contribution of each CANUCS field to the total SFRD. We see that in the $z=2,3,5\ \rm and\ 6$ bins individual fields dominate the SFRD with $>40\%$ contributions. This finding reinforces the need for wider area surveys to accurately trace the contribution of rare starburst galaxies and overdensities to the SFRD given individual sources can significantly impact the measurement \citep{rodighiero11, williams24, sun25}. The ability of the medium-band photometry to identify and characterize such overdensities as observed here thanks to the precise photometric redshifts opens the exciting possibility of incorporating environmental dependencies into future luminosity function or SFRD work \citep[see][for an application with NIRCam grism]{lin25}. This is doubly important due to the tendency of dusty starburst galaxies to reside in overdensities \citep[e.g.,][]{daddi09, calvi23, bhangal25}.

\section{Summary and Conclusion}
\label{sec:Conclusion}

We have presented the first self-consistent measurement of the \halpha\ luminosity function spanning cosmic noon to the epoch of reionization and the associated star formation rate density. The sample of 4101 galaxies is observed in the JWST in Technicolor, CANUCS, and JUMPS surveys with up to 20 \jwst\ NIRCam filters and spans $1.3<z<6.6$. We summarize our main findings as follows:

\begin{itemize}
    \item We demonstrate that \jwst\ medium-band photometry enables photometric redshifts precise and accurate enough to reliably infer emission line fluxes from medium-band flux excess. This enables measurements of SFRs from \halpha\ over a wide range of SFR and $z$ ($1<SFR<1000 \rm\ M_{\odot}\ yr^{-1}$ at $z<2.25$). 
\end{itemize}

\begin{itemize}
    \item We measure the \halpha\ luminosity function in nine redshift bins using a self-consistent dataset and framework. We find good agreement with previous measurements, with the largest discrepancy being a higher observed number density than current NIRCam grism measurements at z=6, potentially driven by an overdensity in the A370 flanking field. 
\end{itemize}

\begin{itemize}
    \item  The luminosity function normalization $\phi^*$ increases with cosmic time, while we observe no significant evolution in the faint-end slope $\alpha$ which maintains values of $\sim-1.6$. The relatively small survey area limits our ability to robustly constrain the bright end and hence the characteristic luminosity $L^*$, though we observe an increase at $z>4$.  
\end{itemize}

\begin{itemize}
    \item  We convert our dust-corrected luminosity functions to star formation rate functions and in general find good agreement with previous determinations from \halpha\, UV, and IR measurements at $z>2.25$. We do observe an elevated high-SFR tail at $z<2.25$ compared to previous \halpha\ measurements, which we attribute to a combination of a different prescription for dust correction and real effects of cosmic variance. 
\end{itemize}

\begin{itemize}
    \item  We report the integrated star formation rate density in both observed and dust-corrected forms across the full redshift range. The comparison of the two suggests obscuration fractions of $\sim 80\%$ at $z=2$, with a decline toward 50\% at $z=4$ where it plateaus up to our highest redshift bin of $z=6$. Additionally, we observe a potential discrepancy with previous measurements, finding values up to $0.2-0.3$ dex higher after correcting for dust. While we consider the result tentative, if correct it would lead to tension with measurements of the stellar mass density, requiring a significant revision of our view of stellar mass growth at cosmic noon.
\end{itemize}

\begin{itemize}
    \item  Our precise photometric redshifts allow us to accurately diagnose overdensities over the full redshift range. We find that two overdensities (in the MACS0417 and Abell 370 flanking fields) significantly affect our star formation rate functions and star formation rate density measurements. 
\end{itemize}

The deep imaging provided in these surveys provides excellent sampling of the faint end of the luminosity function, though the relatively small area limits our interpretation of the evolution of the bright end. Measurements from larger medium-band surveys such as Medium-band Imaging with NIRCam to Explore ReVolutionary Astrophysics \citep[MINERVA][]{muzzin25} will fill this gap (Martis et al. in prep). 

\begin{acknowledgements}
NM, MB, GR, GF, JJ, and VM acknowledge support from the ERC Grant FIRSTLIGHT and Slovenian national research agency ARRS through grants N1-0238 and P1-0188. MB  acknowledges support from the program HST-GO-16667, provided through a grant from the STScI under NASA contract NAS5-26555. Based on observations with the NASA/ESA/CSA James Webb Space Telescope obtained from the Data Archive at the Space Telescope Science Institute, which is operated by the Association of Universities for Research in Astronomy, Incorporated, under NASA contract NAS5-03127. DM acknowledges generous support from the Leonard and Jane Holmes Bernstein Professorship in Evolutionary Science. Support for programs JWST-GO-03362 and JWST-GO-05890, provided through a grant from the STScI under NASA contract NAS5-03127, is acknowledged. This research was supported by grant 18JWST-GTO1 and 23JWGO2A13 from the Canadian Space Agency (CSA), and funding from the Natural Sciences and Engineering Research Council of Canada (NSERC). This research used the Canadian Advanced Network For Astronomy Research (CANFAR) operated in partnership by the Canadian Astronomy Data Centre and The Digital Research Alliance of Canada with support from the National Research Council of Canada the Canadian Space Agency, CANARIE and the Canadian Foundation for Innovation. The Cosmic Dawn Center (DAWN) is funded by the Danish National Research Foundation under grant No. 140. The MAST DOI for CANUCS is \href{https://doi.org/10.17909/ph4n-6n76}{doi:10.17909/ph4n-6n76}. 
\end{acknowledgements}

\paragraph{Facilities.} \textit{HST}(ACS), \jwst\

\paragraph{Software.} astropy \citep{astropy13},  
          photutils \citep{bradley22}

\bibliographystyle{aa}
\bibliography{refs_AA}

\appendix
\section{}
\label{sec:appendix}

In the main body of this work, we have presented \halpha\ luminosity functions in evenly-spaced redshift bins. Here we present the luminosity functions for the redshift ranges corresponding to each medium-band filter individually to allow for direct comparison for other studies using \jwst\ observational data. 

\begin{figure*}[h]
\centering
\includegraphics[width=\textwidth]{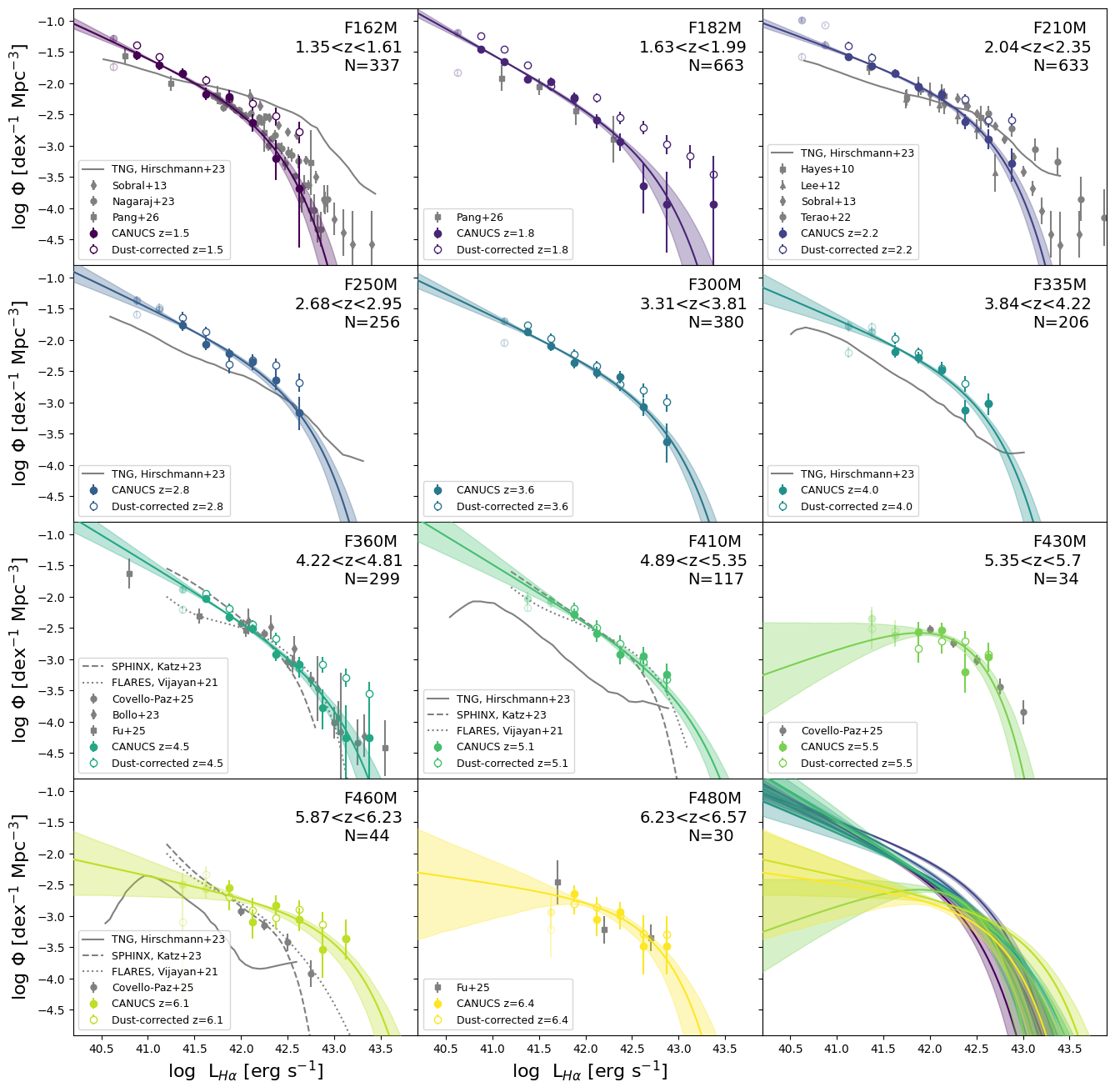}
\caption{\halpha\ luminosity function (colored points) and Schechter fits (colored curves) in each redshift bin corresponding to 11 medium-band filters. For comparison we show a collection of observations including ground-based narrow-band surveys \citep{sobral13}, \textit{Spitzer} photometric measurements from \citet{bollo23}, and \jwst\ NIRCam grism measurements from \citet{sun25, covelo-paz25, fu25, lin25}. We also show predictions from the Illustris TNG cosmological simulation presented in \citet{shen20} (gray curves).}
\label{fig:LF_filters}
\end{figure*}

\begin{table*}
\centering
\begin{tabular}{lccrccc}
\toprule
$z_{mid}$ & $z_{range}$ & $Filter$ & $N$ & $\log L^*$ & $\log \phi^*$ & $\alpha$ \\
\midrule
1.48 & $1.35 < z < 1.61$ & F162M & 337 & $42.26^{+0.16}_{-0.14}$ & $-2.81^{+0.20}_{-0.23}$ & $-1.68^{+0.11}_{-0.10}$ \\
1.81 & $1.63 < z < 1.99$ & F182M & 663 & $42.64^{+0.25}_{-0.24}$ & $-3.24^{+0.30}_{-0.33}$ & $-1.82^{+0.08}_{-0.07}$ \\
2.19 & $2.04 < z < 2.35$ & F210M & 633 & $42.53^{+0.11}_{-0.09}$ & $-2.67^{+0.11}_{-0.14}$ & $-1.55^{+0.06}_{-0.07}$ \\
2.81 & $2.68 < z < 2.95$ & F250M & 256 & $42.52^{+0.15}_{-0.14}$ & $-2.91^{+0.22}_{-0.26}$ & $-1.71^{+0.13}_{-0.14}$ \\
3.56 & $3.31 < z < 3.81$ & F300M & 380 & $42.75^{+0.13}_{-0.13}$ & $-3.21^{+0.19}_{-0.21}$ & $-1.70^{+0.09}_{-0.09}$ \\
4.03 & $3.84 < z < 4.22$ & F335M & 206 & $42.46^{+0.14}_{-0.16}$ & $-2.94^{+0.25}_{-0.26}$ & $-1.63^{+0.18}_{-0.17}$ \\
4.51 & $4.22 < z < 4.81$ & F360M & 299 & $42.89^{+0.25}_{-0.22}$ & $-3.53^{+0.35}_{-0.39}$ & $-1.90^{+0.15}_{-0.13}$ \\
5.12 & $4.89 < z < 5.35$ & F410M & 117 & $43.00^{+0.23}_{-0.20}$ & $-3.65^{+0.37}_{-0.46}$ & $-1.90^{+0.21}_{-0.21}$ \\
5.53 & $5.35 < z < 5.7$ & F430M & 34 & $42.15^{+0.19}_{-0.14}$ & $-2.56^{+0.08}_{-0.18}$ & $-0.46^{+0.46}_{-0.50}$ \\
6.05 & $5.87 < z < 6.23$ & F460M & 44 & $43.00^{+0.22}_{-0.24}$ & $-3.37^{+0.29}_{-0.37}$ & $-1.33^{+0.31}_{-0.24}$ \\
6.40 & $6.23 < z < 6.57$ & F480M & 30 & $42.60^{+0.33}_{-0.32}$ & $-3.18^{+0.36}_{-0.50}$ & $-1.21^{+0.65}_{-0.42}$ \\
\bottomrule
\end{tabular}
\caption{Schechter parameters for the observed H$\alpha$ luminosity function measured with each NIRCam medium-band filter individually. $z_{mid}$ is the median redshift of the observable window for \halpha\ in each filter. $N$ is the number of sources included in the Schechter fit, which uses only sources with luminosity above the $>50\%$ completeness level (excludes sources in the transparent bins in Figure \ref{fig:LF_filters}).}
\label{tab:filterbins}
\end{table*}

\end{document}